\documentclass[aps,prd,twocolumn,superscriptaddress,nofootinbib]{revtex4}
\usepackage{bm}
\usepackage{epsfig}

\begin{document}

\title{Probing low-$\bm{x}$ QCD with cosmic neutrinos at the 
       Pierre Auger Observatory}

\author{Luis A.~Anchordoqui}
\affiliation{Department of Physics,
             Northeastern University, Boston, MA 02115, USA}
\affiliation{Department of Physics, University of Wisconsin-Milwaukee, 
             P.O. Box 413, Milwaukee, WI 53201, USA}
\author{Amanda M. Cooper-Sarkar}
\affiliation{Particle Physics, Denys Wilkinson Laboratory, 
             University of Oxford, Keble Road, Oxford, OX1 3RH, UK}
\author{Dan Hooper} 
\affiliation{Particle Astrophysics Center,
             Fermilab, P.O. Box 500, Batavia, IL 60510, USA}
\author{Subir Sarkar}
\affiliation{Rudolf Peierls Centre for Theoretical Physics, 
             University of Oxford, 1 Keble Road, Oxford, OX1 3NP, UK}

\date{\today}

\begin{abstract}
The sources of the observed ultra-high energy cosmic rays must also
generate ultra-high energy neutrinos. Deep inelastic scattering of
these neutrinos with nucleons on Earth probe center-of-mass energies
$\sqrt{s} \sim 100$~TeV, well beyond those attainable at terrestrial
colliders. By comparing the rates for two classes of observable
events, any departure from the benchmark (unscreened perturbative QCD)
neutrino-nucleon cross-section can be constrained. Using the projected
sensitivity of the Pierre Auger Observatory to quasi-horizontal
showers and Earth-skimming tau neutrinos, we show that a `Super-Auger'
detector can thus provide an unique probe of strong interaction
dynamics.
\end{abstract}



\maketitle

A new window on the Universe is expected to open soon through the
observation of ultra-high energy cosmic neutrinos by detectors such as
the Pierre Auger Observatory~\cite{Abraham:2004dt} and
IceCube~\cite{Achterberg:2006md}. The study of their interactions in
the Earth's atmosphere and crust is particularly interesting as this
provides a probe of QCD in the kinematic region of very small values
of Bjorken-$x$: $x <10^{-3}$, where conventional calculations done in
the DGLAP framework become inadequate.

The surprising discovery at HERA that the gluon distribution function
rises sharply with decreasing $x$ for high virtuality of the exchanged
gauge boson $Q^2$, implies a strong increase in the neutrino-nucleon
total cross-section. This would however violate unitarity (the
Froissart bound) if continued indefinitely, hence there {\em must} be
a departure from this behavior at very low values of~$x$.

Various possibilities have been entertained in this context (see
e.g.,~\cite{Reno:2004cx} for a recent review) which predict a slower
rise of the cross-section than the usual DGLAP evolution based
calculation~\cite{Gandhi:1998ri} commonly adopted for estimating the
effective aperture of detectors, e.g. of Auger, which is sensitive to
neutrinos of energy $\agt 10^8$~GeV~\cite{Capelle:1998zz}. Thus, a
measurement of the neutrino-nucleon cross-section at such energies,
even if relatively crude, would nonetheless be of great interest.  For
example one can directly test models where saturation effects cause
the gluon distribution function to freeze or even {\em decrease} with
$x$ below some threshold, $x \alt 10^{-5}$. The former implies a
$\nu-N$ cross-section which stays about a factor of 2 below the
standard expectation for $E_\nu \sim 10^8 - 10^{10}$~GeV, where Auger
is most sensitive. In the latter case the cross-section remains
sensibly constant with increasing energy above $\sim 10^9$~GeV, so has
a value 10 times below the usual expectation at the upper end of this
range.

One might wonder how such a measurement can be done given the large
uncertainties in the expected fluxes of (the yet to be detected!)
cosmic neutrinos. The opportunity arises because of the discovery that
the muon and tau neutrinos are maximally mixed. Hence cosmic neutrino
beams, generally expected to be $\nu_\mu$'s and $\nu_e$'s from the
decays of pions, kaons and perhaps heavy flavours, will necessarily
contain a substantial component of $\nu_\tau$'s by the time they reach
Earth~\cite{Learned:1994wg}. It has been
noted~\cite{Fargion:2000iz,Bertou:2001vm,Feng:2001ue} that
Earth-skimming $\nu_\tau$'s will generate {\em upward} going air
showers when they interact in the crust. By contrast neutrinos of all
flavours will generate deeply penetrating quasi-horizontal air showers
which are distinctive in having an electromagnetic component unlike
hadron-initiated air showers at such large
inclinations~\cite{Zas:2005zz}. The crucial
observation~\cite{Kusenko:2001gj} is that while the rate of
quasi-horizontal $\nu$-showers is proportional to the $\nu-N$
cross-section, the rate of detectable Earth-skimming $\nu_\tau$'s is
not. This is because of several effects which come into play in the
latter case. For example, a decrease in the cross-section will
increase the number of $\tau$'s produced in a region close enough to
Earth's surface that are likely to escape. However, the resulting
$\tau$'s emerge preferentially at angles outside the acceptance
typical of surface arrays. Consequently, the number of detected
$\tau$'s is relatively independent of the cross-section. All in all,
the {\em ratio} of the two classes of events provides a measure of the
absolute $\nu-N$ cross-section, even when there are large
uncertainties in the incoming cosmic flux.

For example although the (so far unknown) sources of the observed
ultra-high energy cosmic rays {\em must} also be sources of ultra-high
energy neutrinos, it is difficult to calculate the expected neutrino
flux in terms of the observed cosmic ray flux, given our ignorance of
the opacity of the sources. The usual benchmark here is the so-called
(all flavours) Waxman-Bahcall flux~\cite{Waxman:1998yy}
\begin{equation}
 \phi^\nu_{\rm WB} \simeq 4 \times 10^{-8} 
                   \left(E_\nu/\mathrm{GeV}\right)^2  
 \mathrm{GeV}\,\mathrm{cm}^{-2}\,\mathrm{s}^{-1}\,\mathrm{sr}^{-1},
\label{wb}
\end{equation}
for $10^3 \lesssim E_\nu/\mathrm{GeV} \lesssim 10^{11}$, derived
assuming that the sources are `transparent' and that 60\% of the proton
energy is converted to pions. We will use this flux to estimate the
event rates for various models of the neutrino scattering cross-section.

The cross-section for charged current (CC) $\nu-N$ scattering
is~\cite{Cooper-Sarkar:1997jk}
\begin{equation}
 \sigma_{\rm pQCD} =\int_0^1 \mathrm{d}x \int_0^{xs} \mathrm{d}Q^2 
 {\mathrm{d}^2 \sigma^{\nu N} \over \mathrm{d}x \mathrm{d}Q^2}\, ,
\end{equation}
where 
\begin{eqnarray}
 {\mathrm{d}^2\sigma^{\nu N} \over \mathrm{d}x \mathrm{d}Q^2} & = & 
 {G_\mathrm{F}^2 \over 2\pi x}
 \bigg({m^2_W \over Q^2 + m^2_W}\bigg)^2 
 \bigg[Y_+\, F_2 (x, Q^2) \nonumber \\
 & - & y\, F_{\rm L}(x, Q^2) + Y_-\, xF_3(x, Q^2)\bigg]
\end{eqnarray}
is the differential cross-section given in terms of the structure
functions $F_2,$ $F_{\rm L}$ and $xF_3$, and $Y_+ = 1 + (1-y)^2$, $Y_-
= 1 - (1-y)^2$ with $y = Q^2/sx$ ($s = 2 E_\nu m_N$ is the
center-of-mass (c.m.) energy, $G_{\rm F}$ is the Fermi constant and
$m_W$ is the $W$-boson mass). For simplicity, we are considering only
CC interactions, as neutral current (NC) interactions are
subdominant. At leading order (LO) in perturbative QCD, the structure
functions are given in terms of parton distributions as $F_2 = \sum_i
x [q_i(x,Q^2) + \bar{q}_i(x, Q^2)],$ $xF_3 = \sum_i x [q_i(x, Q^2) +
\bar{q}_i(x, Q^2)]$ and $F_{\rm L} = 0$.  However at NLO these
relationships involve further QCD-calculable coefficient functions and
contributions from $F_{\rm L}$ can no longer be neglected. Parton
distribution functions (PDFs) are determined in fits to deep inelastic
scattering (DIS) data by the following procedure. The parton
distribution functions are parameterised at some initial scale $Q_0
\sim 1$~GeV and then evolved, using the next-to-leading order (NLO)
DGLAP equations, to higher values of $Q^2$; they are then convoluted
with QCD-calculable coefficient functions to give NLO predictions for
the structure functions, which are then fitted to the DIS
data~\cite{Cooper-Sarkar:1997jk}. Such fits have been made by several
different groups~\cite{Martin:2001es,Pumplin:2002vw,Chekanov:2002pv};
recent analyses have included estimates of the uncertainties on the
PDFs coming from experimental uncertainties.

As $Q^2$ increases, the parton distribution functions (particularly of
the gluon) grow due to QCD evolution, so that the neutrino
cross-section will also grow until the propagator cuts off the growth
at $Q^2 \sim m_W^2$. Hence the typical $x$ value probed is $x \sim
m_W^2/ 2 m_N E_\nu$~\cite{Frichter:1994mx}. For neutrino energies
$E_\nu \sim 10^8 - 10^{10}$~GeV this translates into small $x$ values
of $10^{-4} - 10^{-6}$ at $Q^2 \sim 10^4$~GeV$^2.$ HERA measurements
do extend down to $x \sim 10^{-6},$ but only at $Q^2 < 0.1$~GeV$^2$,
while for $Q^2 \sim m_W^2$ the LHC will probe $5\times 10^{-4} < x <
5\times 10^{-2}$. To probe down to the same kinematic region as Auger
would require a hadron collider with c.m. energy exceeding $10^3$~TeV.

\begin{figure}
\setlength{\epsfxsize}{0.98\hsize}\centerline{\epsfbox{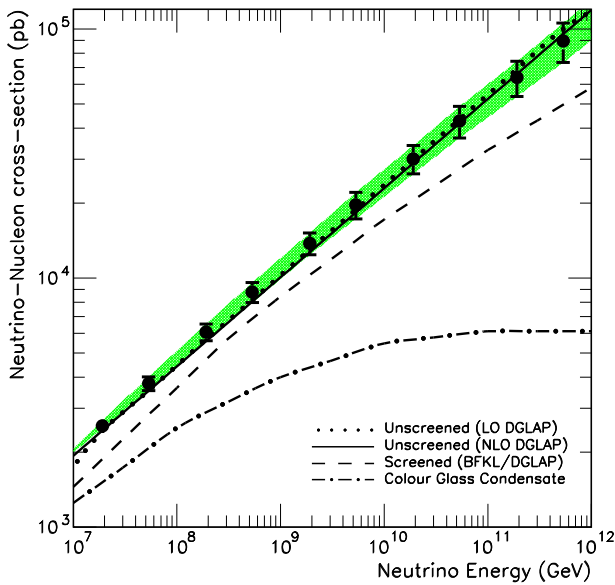}}
 \caption{Predicted cross-sections for neutrino-nucleon scattering at
 high energies. The line with its $1 \sigma$ error band is
 the fit (Eq.\protect\ref{nlo}) to our calculated 
 $\sigma_{\rm uns}^{\rm NLO} (E_\nu)$ (points with error bars). 
 For comparison we show $\sigma_{\rm uns}^{\rm LO} (E_\nu)$ (dotted line)
 \protect\cite{Gandhi:1998ri}, $\sigma_{\rm scr}^{\rm KK} (E_\nu)$ (dashed
 line) \protect\cite{Kutak:2003bd}, and $\sigma_{\rm scr}^{\rm HJ} (E_\nu)$
 (dot-dashed line) \protect\cite{Henley:2005ms}.}
\label{fig:sigma}
\end{figure}

At small $x$ and high $Q^2$ the $\nu-N$ cross-section is dominated by
sea quarks produced by gluon splitting $g \to q \bar q.$ In this
kinematic region, the parametrisation of the gluon momentum distribution is
approximately: $xg(x,Q_0^2) \propto x^{-\lambda}$ for $x \ll 1$,
where $\lambda \simeq 0.3-0.4$. The resulting CC $\nu-N$ cross-section
was originally calculated at leading order using 1996 parton distribution
functions and parametrised as: $\sigma_\mathrm{unscr}^\mathrm{LO} =
5.53 \, (E_\nu/\mathrm{GeV})^{0.363}~{\rm pb}$ for $10^7 \leq E_\nu
\leq 10^{12}$~GeV~\cite{Gandhi:1998ri}. This is the benchmark {\em
unscreened} cross-section widely used to evaluate sensitivities of
ultra-high energy cosmic neutrino detectors. We show this in
Fig.~\ref{fig:sigma} along with our updated calculation using a
modern PDF analysis \cite{Chekanov:2002pv} which included the final
data from the first phase of HERA running (1993--2000); this data is
{\em essential} to obtain information on the PDFs at low-$x$. Our NLO 
analysis includes corrections for heavy flavours, and, most
importantly, a full treatment of experimental uncertainties 
which were not considered in~\cite{Gandhi:1998ri}. 
We also show in Fig.~\ref{fig:sigma} a simple power-law fit to our 
improved CC $\nu-N$ cross-section:
\begin{equation}
\sigma_\mathrm{unscr}^\mathrm{NLO} = 
6.04 \pm 0.40 \, (E_\nu/\mathrm{GeV})^{0.358 \pm 0.005}~\mathrm{pb}\ ,
\label{nlo}
\end{equation}
for $10^7 \leq E_\nu \leq 10^{12}$~GeV. (The tabulated cross-section
and other details are provided elsewhere \cite{mandy}.)

However, when $x$ is small two further considerations are relevant.
First, when $x$ is sufficiently small that $\alpha_\mathrm{s}\,\ln
(1/x) \sim 1$, it is also necessary to resum these large logarithms,
using the BFKL formalism. Second, the gluon density at small-$x$ is
very high and both the DGLAP and the BFKL formalisms neglect
non-linear screening effects due to gluon recombination. Such effects
would tame the rise of the gluon distribution function at small $x$
and may even lead to saturation. An efficient way of modelling this is
the color dipole framework in which DIS at low $x$ is viewed as the
interaction of the $q \bar q$ dipole to which the gauge bosons
fluctuate. An unified BFKL/DGLAP calculation supplemented by screening
effects, as well as nuclear shadowing (following the calculation in
\cite{CastroPena:2000sx} for $A = 12$), predicts a decrease of the
cross-section $\sigma_\mathrm{scr}^{\rm KK}$ by a factor of 
$\sim 1.2-2$ in the relevant energy range~\cite{Kutak:2003bd}.  An
alternative recent approach uses the colour glass condensate
formalism~\cite{Iancu:2003xm}; this predicts a similar suppression
when a dipole model~\cite{Kharzeev:2004yx} which fits data from RHIC is
used~\cite{Henley:2005ms}. The predicted cross-section
$\sigma_\mathrm{scr}^\mathrm{HJ}$ is even lower if a different dipole
model developed to fit the HERA data~\cite{Bartels:2002cj} is used and
the gluon distribution is assumed (more speculatively) to {\em
decrease} for $x < 10^{-5}$~\cite{Henley:2005ms}. As seen in
Fig.~\ref{fig:sigma} this is a factor of $\sim 2-20$ below the unscreened
cross-section in the relevant energy range. Other possibilities for
the high energy $\nu-N$ cross-section have been
discussed~\cite{Jalilian-Marian:2003wf}; an exciting development
formulates DIS using gauge/string duality and provides new insights
into low-$x$ dynamics \cite{Polchinski:2001tt}.

The Pierre Auger Observatory is the largest cosmic ray detector in the
world~\cite{Abraham:2004dt} and when complete will occupy two sites
--- one in the Southern and one in the Northern
hemisphere. Construction of Auger South on a plateau in Western
Argentina is well advanced and it is operating in hybrid mode
employing fluorescence detectors overlooking a $\sim 3000$ km$^2$
ground array of water \v{C}erenkov detectors. In addition to studying
the highest energy cosmic rays, Auger is also capable of observing
ultra-high energy cosmic neutrinos
\cite{Capelle:1998zz,Bertou:2001vm}.

\begin{table}
\caption{Neutrino events per year at Auger for different models of the
 $\nu - N$ cross-section, adopting two benchmark cosmic fluxes and
 shower energy threshold of $10^8 (10^9)$~GeV.}

\begin{tabular}{|c@{}|c|c@{}|c|c|}
  \hline
  \hline
  Model~ & \multicolumn{2}{@{}c}{Waxman-Bahcall} & \multicolumn{2}{@{}c|}{Cosmogenic} \\
  \hline
  \cline{1-3} \cline{4-5}
  &~~${\cal N}_\mathrm{QH}$~~&~~${\cal N}_\mathrm{ES}$~~&~~${\cal N}_\mathrm{QH}$~~&~~${\cal N}_\mathrm{ES}$~~ \\
  \hline
  \hline
  $\sigma_\mathrm{unscr}^{\rm LO}$ & 0.15\phantom{0}~(0.092)~& 3.0~(0.62)~~& 0.061~(0.039)~& 1.2~(0.35)~\\
  \hline
  $\sigma_\mathrm{unscr}^{\rm NLO}$ & 0.14\phantom{0}~(0.080)~& 3.0~(0.61)~~& 0.057~(0.036) & 1.2~(0.34)~\\
  \hline
  $\sigma_\mathrm{scr}^{\rm  KK}$ & 0.10\phantom{0}~(0.057)~& 2.7~(0.54)~~& 0.042~(0.027) & 1.1~(0.31)~\\
  \hline
  $\sigma_\mathrm{scr}^{\rm  HJ}$ & 0.048~(0.022)~& 1.8~(0.32)~~& 0.018~(0.010) & 0.7~(0.18)~\\
  \hline
  \hline
\end{tabular}
\label{table}
\end{table}

Given an isotropic flux of neutrinos $\phi^\nu (E_\nu)$, the rate of
quasi-horizontal (QH) showers expected to be observed at Auger is
proportional to the $\nu-N$ cross-section:
\begin{equation}
 {\cal N}_{\rm QH} \propto \int \mathrm{d}E_{\rm sh} \,\,\sigma_{\nu N} (E_\nu)
 \,\, A_{\rm QH} (E_{\rm sh}, \theta_{\rm z}) \,\, \phi^\nu (E_\nu)\, .
\end{equation}
Here $A_{\rm QH}(E_{\rm sh}, \theta_{\rm z})$ is the Auger acceptance,
which depends on the zenith angle $\theta_{\rm
z}$~\cite{Anchordoqui:2005ey}, and $E_{\rm sh} = y E_\nu$ for $\nu_\mu$
and $\nu_\tau$, whereas $E_{\rm sh} = E_\nu$ for $\nu_e$.  The
expected event rate for the WB flux (Eq.\ref{wb}) is given in
Table~\ref{table}.

The situation is different for showers initiated by $\tau$'s created
by CC interactions of Earth-skimming (ES) $\nu_\tau$'s. To a first
approximation the number of such events is
\begin{equation}
{\cal N}_\mathrm{ES} \propto \int \mathrm{d}E_\mathrm{sh}\
 \mathrm{d}\cos\theta\ \mathrm{d}\phi\ P(\theta, \phi) A_\mathrm{ES}
 (E_\mathrm{sh}, \theta) \phi^\nu (E_\nu)\ ,
\end{equation}
where
\begin{equation}
P (\theta, \phi) = \int_0^\ell \frac{\mathrm{d}z}{l_\nu^\mathrm{CC}}
 \mathrm{e}^{-z/l_\nu^\mathrm{tot}}\ \Theta \left[z - (\ell - l_\tau) \right]
\label{P}
\end{equation}
is the probability for a $\nu_\tau$ with incident nadir angle $\theta$
and azimuthal angle $\phi$ to emerge as a detectable $\tau$. Here
$l_\tau \sim 10$~km is the typical $\tau$ path
length~\cite{Feng:2001ue}, $\ell = 2 R_{\oplus} \cos\theta$ is the
chord length of the intersection of the neutrino trajectory with the
Earth (of radius $R_{\oplus} \approx 6371$~km), $l_\nu^\mathrm{CC}$
and $l_\nu^\mathrm{tot}$ are the CC and total neutrino mean free
paths, respectively, and $A_\mathrm{ES} (E_\mathrm{sh}, \theta)$ is
the experimental acceptance which has a strong dependence on the
angle, since the surface detector array can only see events within a
few degrees of the horizon~\cite{Zas:2005zz}. In fact, the analytic
expression above is an oversimplification; it does not allow for
$\tau$ regeneration in the Earth and the $\tau$ path length is not
really a step-function. To take such details into account, we have
carried out a simple Monte Carlo simulation assuming the NC
cross-section to be 40\% of the CC cross-section for all models. The
resulting ES event rates are given in Table~\ref{table} --- we find
good agreement with results from a sophisticated Monte
Carlo that models the environment of Auger and its acceptance 
accurately \cite{markus}.

\begin{figure}
\setlength{\epsfxsize}{0.98\hsize}\centerline{\epsfbox{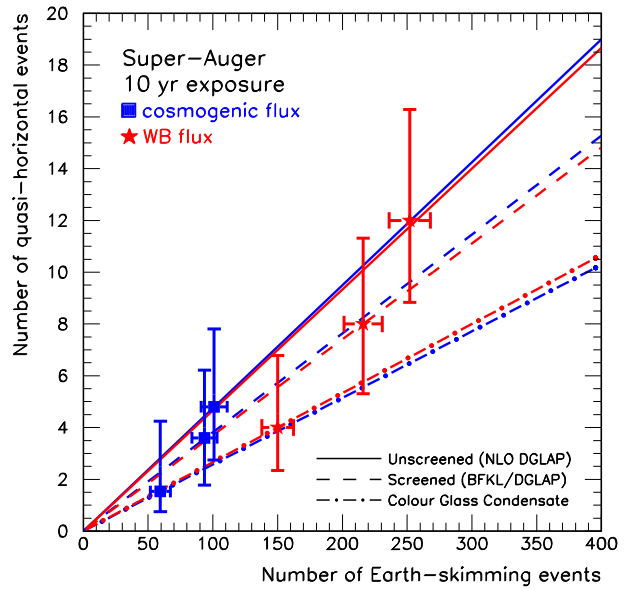}}
\caption{The expected number of Earth-skimming and quasi-horizontal
 neutrino events above $10^8$~GeV for different models of the $\nu-N$
 cross-section; for each model, one of the lines assumes the cosmogenic
 spectrum and the other line the Waxman-Bahcall spectrum. We show as
 squares and stars respectively, the corresponding hypothetical
 measurements (with $1 \sigma$ statistical errors) that could be made
 in 10~yr with an Auger-like detector scaled up to 10000 square miles.}
\label{fig:sensitivity}
\end{figure}

To evaluate the sensitivity to the assumed spectral index we also
consider the ``guaranteed'' cosmogenic neutrino flux which has a
peaked distribution in the energy range of
interest~\cite{Engel:2001hd}.  As shown in Table~\ref{table}, if we
consider events with $E_{\rm sh}^{\rm th} > 10^8$~GeV, the change in
the spectrum produces a variation in the ratio ${\cal N}_{\rm
QH}/{\cal N}_{\rm ES}$ of less than 5\%. We have also verified that a
more steeply falling flux $\propto E^{-2.54}$ (with the same $E_{\rm
sh}^{\rm th}$) causes a change in ${\cal N}_{\rm QH}/{\cal N}_{\rm
ES}$ by about 10\% (e.g, for $\sigma_\mathrm{unscr}^{\rm NLO}$, ${\cal
N}_{\rm QH} = 0.69~{\rm yr}^{-1}$ and $ {\cal N}_{\rm ES}= 12~{\rm
yr}^{-1}$). Such a flux is expected~\cite{Ahlers:2005sn} if
extragalactic cosmic rays from `transparent' sources begin dominating
the observed spectrum at $10^{9.6}$~GeV \cite{Bergman:2004bk} rather
than at $\sim 10^{10.5}$~GeV as is usually assumed (see Fig.~5
in~\cite{Anchordoqui:2005ey} for a comparison of these fluxes). Thus
we conclude that the ratio ${\cal N}_{\rm QH}/{\cal N}_{\rm ES}$
provides a {\em robust} estimate of the $\nu-N$
cross-section~\cite{fluorescence}.

Now we discuss the precision with which such a measurement can be
made.  In Fig.~\ref{fig:sensitivity} we show the numbers of QH and ES
events expected for different cross-section predictions and flux
expectations. The data points are hypothetical measurements made over
10 years with a proposed `Super Auger' array of area 10000 square
miles (25600 km$^2$). Note that by observing just a dozen QH events,
we can begin to distinguish between the theoretical models, however
from Table~\ref{table} we see that the number of events for the WB
flux is an order of magnitude smaller in Auger even after 10
years. Hence we are led to entertain the idea of scaling up the array
by an order of magnitude, perhaps by using radio detection methods
\cite{jim}.  Another possibility is to use satellite-borne
fluorescence detectors such as EUSO and OWL which may attain the
required order of magnitude increase in sensitivity
\cite{Palomares-Ruiz:2005xw}.

The cosmic neutrino flux can be much higher than the conservative
benchmark values we have adopted above.  Data from
HiRes~\cite{Bergman:2004bk} suggest that it may
be~\cite{Ahlers:2005sn} just below the current experimental bound from
AMANDA-B10~\cite{Ackermann:2005sb}. The expected event rates in Auger
itself would then be high enough to test for any suppression of the
UHE $\nu-N$ cross-section. Since the spectrum in this case is softer
than the WB flux, IceCube~\cite{Achterberg:2006md}, with its lower
energy threshold, should test this model very soon. Indeed IceCube has
the sensitivity to see the benchmark WB flux within a few years. Thus
there is an emerging synergy between large cosmic ray arrays and
cosmic neutrino detectors which will soon establish if there is a
realistic possibility for exploring fundamental physics using Nature's
own high energy beams.  Given that there is no conceivable terrestrial
accelerator which can attain such energies, we believe that the
construction of Super Auger should be considered seriously; the
detection techniques are well developed and the resources required are
no larger than for a contemporary collider experiment.

\smallskip
\noindent 
LAA is partially supported by NSF Grant No. PHY-0457004. DH is
supported by the DOE and by NASA Grant NAG5-10842. SS acknowledges a
PPARC Senior Research Fellowship (PPA/C506205/1). We wish to thank Jim
Cronin and Alan Watson for discussions and encouragement.

\end{document}